\PassOptionsToPackage{unicode}{hyperref}
\PassOptionsToPackage{hyphens}{url}
\PassOptionsToPackage{dvipsnames,svgnames,x11names}{xcolor}
\documentclass[
  12pt,
]{article}
\usepackage{xcolor}
\usepackage{amsmath,amssymb}
\usepackage{iftex}
\ifPDFTeX
  \usepackage[T1]{fontenc}
  \usepackage[utf8]{inputenc}
  \usepackage{textcomp} 
\else 
  \ifXeTeX
    \usepackage{mathspec} 
  \else
    \usepackage{unicode-math} 
  \fi
  \defaultfontfeatures{Scale=MatchLowercase}
  \defaultfontfeatures[\rmfamily]{Ligatures=TeX,Scale=1}
\fi
\usepackage{lmodern}
\ifPDFTeX\else
\fi
\IfFileExists{upquote.sty}{\usepackage{upquote}}{}
\IfFileExists{microtype.sty}{
  \usepackage[]{microtype}
  \UseMicrotypeSet[protrusion]{basicmath} 
}{}
\makeatletter
\@ifundefined{KOMAClassName}{
  \IfFileExists{parskip.sty}{%
    \usepackage{parskip}
  }{
    \setlength{\parindent}{0pt}
    \setlength{\parskip}{6pt plus 2pt minus 1pt}}
}{
  \KOMAoptions{parskip=half}}
\makeatother
\makeatletter
\ifx\paragraph\undefined\else
  \let\oldparagraph\paragraph
  \renewcommand{\paragraph}{
    \@ifstar
      \xxxParagraphStar
      \xxxParagraphNoStar
  }
  \newcommand{\xxxParagraphStar}[1]{\oldparagraph*{#1}\mbox{}}
  \newcommand{\xxxParagraphNoStar}[1]{\oldparagraph{#1}\mbox{}}
\fi
\ifx\subparagraph\undefined\else
  \let\oldsubparagraph\subparagraph
  \renewcommand{\subparagraph}{
    \@ifstar
      \xxxSubParagraphStar
      \xxxSubParagraphNoStar
  }
  \newcommand{\xxxSubParagraphStar}[1]{\oldsubparagraph*{#1}\mbox{}}
  \newcommand{\xxxSubParagraphNoStar}[1]{\oldsubparagraph{#1}\mbox{}}
\fi
\makeatother

\usepackage{longtable,booktabs,array}
\usepackage{calc} 
\usepackage{etoolbox}
\makeatletter
\patchcmd\longtable{\par}{\if@noskipsec\mbox{}\fi\par}{}{}
\makeatother
\IfFileExists{footnotehyper.sty}{\usepackage{footnotehyper}}{\usepackage{footnote}}
\makesavenoteenv{longtable}
\usepackage{graphicx}
\makeatletter
\newsavebox\pandoc@box
\newcommand*\pandocbounded[1]{
  \sbox\pandoc@box{#1}%
  \Gscale@div\@tempa{\textheight}{\dimexpr\ht\pandoc@box+\dp\pandoc@box\relax}%
  \Gscale@div\@tempb{\linewidth}{\wd\pandoc@box}%
  \ifdim\@tempb\p@<\@tempa\p@\let\@tempa\@tempb\fi
  \ifdim\@tempa\p@<\p@\scalebox{\@tempa}{\usebox\pandoc@box}%
  \else\usebox{\pandoc@box}%
  \fi%
}
\def\fps@figure{htbp}
\makeatother

\usepackage[giveninits=true]{biblatex}
\usepackage[utf8]{inputenc}
\usepackage{amsmath,amssymb,amsfonts,amsthm}
\usepackage{mathfont}
\usepackage{hyperref}
\usepackage[capitalise,noabbrev]{cleveref}
\usepackage[noblocks]{authblk}

\usepackage{setspace}
\usepackage[margin=1in]{geometry}
\makeatletter
\@ifpackageloaded{caption}{}{\usepackage{caption}}
\AtBeginDocument{%
\ifdefined\contentsname
  \renewcommand*\contentsname{Table of contents}
\else
  \newcommand\contentsname{Table of contents}
\fi
\ifdefined\listfigurename
  \renewcommand*\listfigurename{List of Figures}
\else
  \newcommand\listfigurename{List of Figures}
\fi
\ifdefined\listtablename
  \renewcommand*\listtablename{List of Tables}
\else
  \newcommand\listtablename{List of Tables}
\fi
\ifdefined\figurename
  \renewcommand*\figurename{Figure}
\else
  \newcommand\figurename{Figure}
\fi
\ifdefined\tablename
  \renewcommand*\tablename{Table}
\else
  \newcommand\tablename{Table}
\fi
}
\@ifpackageloaded{float}{}{\usepackage{float}}
\floatstyle{ruled}
\@ifundefined{c@chapter}{\newfloat{codelisting}{h}{lop}}{\newfloat{codelisting}{h}{lop}[chapter]}
\floatname{codelisting}{Listing}

\makeatother
\makeatletter
\@ifpackageloaded{caption}{}{\usepackage{caption}}
\@ifpackageloaded{subcaption}{}{\usepackage{subcaption}}
\makeatother
\usepackage{bookmark}
\IfFileExists{xurl.sty}{\usepackage{xurl}}{} 
\hypersetup{
  pdftitle={Stable and practical semi-Markov modelling of intermittently-observed data},
  pdfauthor={Christopher Jackson},
  colorlinks=true,
  linkcolor={blue},
  filecolor={Maroon},
  citecolor={Blue},
  urlcolor={Blue},
  pdfcreator={LaTeX via pandoc}}

\title{Stable and practical semi-Markov modelling of
intermittently-observed data}

\author{
      {Christopher Jackson}~\\
          MRC Biostatistics Unit, University of Cambridge~\\
      \footnotesize
      East Forvie Building, Robinson Way, Cambridge Biomedical Campus,
Cambridge, United Kingdom, CB2 0SR~\\
      
        \href{mailto:chris.jackson@mrc-bsu.cam.ac.uk}{chris.jackson@mrc-bsu.cam.ac.uk}
    \and 
  }

\date{}

\begin{document}
\maketitle

\hypersetup{pdfcreator={Christopher Jackson}}
\newcommand{\btheta}{{\boldsymbol{\theta}}}
\newcommand{\bbeta}{\boldsymbol{\beta}}
\newcommand{\bgamma}{\boldsymbol{\gamma}}
\newcommand{\blambda}{\boldsymbol{\lambda}}
\newcommand{\x}{{\mathbf{x}}}
\newcommand{\y}{{\mathbf{y}}}
\newcommand{\z}{{\mathbf{z}}}
\newcommand{\h}{{\mathbf{h}}}

\begin{abstract}
Multi-state models are commonly used for intermittent observations of a state over time, but these are generally based on the Markov assumption, that transition rates are independent of the time spent in current and previous states.  In a semi-Markov model, the rates can depend on the time spent in the current state, though available methods for this are either restricted to specific state structures or lack general software.  This paper develops the approach of using a ``phase-type" distribution for the sojourn time in a state, which expresses a semi-Markov model as a hidden Markov model, allowing the likelihood to be calculated easily for any state structure.  While this approach involves a proliferation of latent parameters, identifiability can be improved by restricting the phase-type family to one with similar properties to a simpler distribution such as the Gamma or Weibull.  This paper proposes a moment-matching method to obtain this family, making general semi-Markov models for intermittent data accessible in software for the first time.  The method is implemented in a new R package, \texttt{msmbayes}, which implements Bayesian or maximum likelihood estimation for multi-state models with general state structures and covariates.  The software is tested through simulation-based calibration, and an application to cognitive function decline illustrates the use of the method in a typical modelling workflow.
\end{abstract}

\section{Introduction}\label{sec-intro}

Many scientific analyses involve studying how a categorical variable
changes over time. This can be done with a \emph{multi-state model}, a
continuous-time stochastic process on a finite set of states. In
biomedical and other application areas, it is common that the state of
an individual is only observed at a finite set of times, giving
``intermittently-observed'' or ``panel'' data. Unlike multistate models
for conventional time-to-event data \autocite{putter2007tutorial}, times
of entry to and exit from states, and which transitions may have occured
between observations, are unknown in general. Such data are generally
modelled with Markov models, using the maximum likelihood method
introduced by \textcite{kalbfleisch1985analysis} and its various
extensions. The \texttt{msm} R package \autocite{jackson2011multi} is
widely used for this. See, e.g. \textcite{van2016multi} or
\textcite{cook2018multistate} for broad reviews.

In these models, it is particularly challenging to relax the Markov
assumption, that is, to allow an individual's transition rates to depend
on the previous states they have visited and their transition times.
This is inherently difficult with intermittent observations, since this
history is only partially known. A related challenge of multi-state
modelling of intermittent observations is the disconnection between the
model parameters (transition rates in continuous time) and the observed
data. This is because the pathway taken through the states between the
observation times, and the times of the transitions on that pathway, are
unknown in general, so it is hard to determine what parameters might be
feasible to estimate from the data without attempting to fit the model.
It is common (in my experience as the maintainer of the \texttt{msm}
package) for users to fit over-parameterised models, where the
parameters are weakly identifiable. This typically results in failure of
the optimisation algorithm.

Here we focus on semi-Markov models, in which transition rates depend on
the time since entry to the current state. This time is unknown, since
the time of entry is unknown. A range of approaches for semi-Markov
modelling of panel data have been suggested for general transition
structures.

One general approach is based on integrating over the latent transition
times \autocite{aastveit2023new,wei2016semi,kapetanakis2013semi}, though
the computational cost of doing this increases substantially with the
number of unobserved transitions. A nonparametric estimator was also
devised by \textcite{gu2023maximum} , but only for transition structures
without cycles. An alternative approach which avoids explicit
integration, and allows any transition structure, is to iteratively
simulate pathways consistent with the data, as part of a custom MCMC
scheme \autocite{barone2022bayesian} or a stochastic EM algorithm
\autocite{aralis2019stochastic}, though software for this is not
available.

Another approach general to any structure, is the ``phase-type'' model
proposed by \textcite{titman2010semi}. As this uses standard maximum
likelihood rather than a specialised fitting algorithm, it is more
amenable to a software implementation. This approach replaces the
exponentially-distributed sojourn in a state by a sequence of
exponential sojourns in latent ``phases''. This expresses a semi-Markov
model as a hidden Markov model, which allows likelihoods to be evaluated
easily. However it introduces many extra parameters, which can lead to
poor identifiability, particularly for intermittent observations. To
address this problem, \textcite{titman2014estimating} developed a method
that uses the phase-type structure to approximate simpler time-to-event
distributions, specifically the Weibull and Gamma, however the
``variational optimization'' used to determine the approximation relied
on an intensive numerical search and ad-hoc choices of spline functions,
and code was not published.

This paper describes a new computational procedure that makes stable
semi-Markov modelling of intermittent data, with general transition
structures, accessible in general software for the first time. This uses
phase-type families inspired by standard distributions, similar to
\textcite{titman2014estimating}, but instead of using numerical
optimization to obtain an approximating distribution family, a fast
analytic procedure is used. This involves finding the unique phase-type
distribution of a particular family whose first three moments agree with
those of the Gamma or Weibull. The model from that paper is also
extended here to include covariates. The method is implemented in a new
R package, \texttt{msmbayes}, which also allows general multi-state
models for intermittently-observed data to be fitted by either Bayesian
estimation or maximum likelihood. An advantage of a Bayesian approach is
that background information can be used to stabilise computation in the
situation where maximum likelihood fails due to weak identifiability.

Section~\ref{sec-models} gives an overview of the current framework for
multi-state modelling of intermittently-observed data.
Section~\ref{sec-smm} describes phase-type semi-Markov models, their use
to approximate simpler distributions, and the novel method introduced to
implement this approximation. The computational methods and software are
introduced in Section~\ref{sec-comp}. Section~\ref{sec-sbc} presents a
simulation-based calibration study to demonstrate that the software
produces the correct posterior for a wide range of model classes.
Section~\ref{sec-app} demonstrates how the models might be used in a
realistically-complex example based on a dataset measuring transitions
between states of cognitive function and death.

\section{Overview of multi-state models}\label{sec-models}

In a multi-state model, an individual moves between a set of states
\(1,\ldots,R\) in continuous time according to \emph{transition
intensities} (or rates) \(q_{rs}(t, \mathcal{F}_t)\), defined as

\[ lim_{\delta t \rightarrow 0} \frac{P(S(t + \delta t) = s | S(t) = r, \mathcal{F}_t)}{\delta t} \]

where the state at time \(t\) since the beginning of the process is
\(S(t)\). If an individual cannot move to \(s\) immediately on exiting
\(r\), then \(q_{rs}=0\). The \emph{transition intensity matrix} \(Q\)
is defined to have \((r,s)\) entry \(q_{rs}\) for \(r \neq s\), and
\(r\)th diagonal entry \(-\sum_{s \neq r} q_{rs}\).

\(\mathcal{F}_t\) represents the \emph{history} of the process up to
time \(t\), that is, the states previously occupied by the individual
and the times of transition between them. This paper is only concerned
with \emph{Markov} and \emph{semi-Markov} models. In a Markov model, the
transition rates are independent of the history \(\mathcal{F}_t\). In a
time-homogeneous Markov model, the \emph{sojourn distribution} in a
state \(r\) (the time spent there before moving to another state) is
exponential with rate \(\sum_{s \neq r} q_{rs}\).

The transition intensity can be related to covariates
\({\mathbf{x}}(t)\), usually via a proportional intensities model with
\(q_{rs}({\mathbf{x}}(t)) = q_{rs}^{(0)}\exp(\boldsymbol{\beta}^T {\mathbf{x}}(t))\).
If the covariates depend on time \(t\), the model is
\emph{time-inhomogeneous}. A model can be time-inhomogeneous but still
Markov, if covariates depend on time \(t\) since the start of the
process, but not on the times spent in particular states.

An important quantity for fitting and prediction is the \emph{transition
probability matrix} \(P(t)\) which has \((r,s)\) entry
\(p_{rs}(t) = P(S(u+t) = s |
S(u) = r)\). If covariates are assumed to be piecewise-constant in time,
\(P(t)\) can be calculated easily (though relaxations are possible,
\textcite{titman2011flexible}). Over any interval \((u,u+t)\) where the
covariates \({\mathbf{x}}\) are constant, \(q_{rs}({\mathbf{x}})\) is
constant, hence \(P(t) = Exp(tQ({\mathbf{x}}))\), where \(Exp()\) is the
matrix exponential \autocite[see, e.g.][]{cox:miller}.

The data consist of observations of the state \(S_{i,j}\) for
independent individuals \(i\) at times \(t_{i,j}\). The likelihood for
individual \(i\) is calculated as a product (over \(j\)) of the
probabilities of occupying \(S_{i,j+1}\) at \(t_{i,j+1}\) given states
at times \(t_{i,j}\) and earlier. For a Markov model, these
probabilities are independent of the history before \(t_{i,j}\), so the
likelihood follows directly from the definition of \(p_{rs}(t)\), as
\(L({\boldsymbol{\theta}}) = \prod_{i, j} p_{S_{i,j},S_{i,{j+1}}}(t_{i,j+1} - t_{i,j})\),
where the \({\boldsymbol{\theta}}\) include the set of \(q_{rs}\), or
the \(q_{rs}^{(0)}\) and \(\boldsymbol{\beta}\)
\autocite{kalbfleisch1985analysis}. Therefore the likelihood can be
calculated easily if the covariates are assumed to be constant between
observations. There are simple extensions of this likelihood to cases
where the entry time to some states is known exactly, or where the state
is known only to be within a particular subset of states
\autocite{kay1986markov}.

A particularly important extension is the (continuous time) \emph{hidden
Markov model}, where observations of an outcome are assumed to be
generated independently conditionally on the state of a Markov
multi-state model. These models have been used for data from medical
diagnostic tests that are subject to error, so that the observed data at
time \(t_{i,j}\) is a state \(O_{i,j}\) that may be different from the
true state \(S_{i,j}\) \autocite{jackson2003multistate}. Hidden Markov
models can also be used as a mechanism for implementing semi-Markov
models, as we show in the following section. The joint likelihood
function for transition intensities and outcome probabilities can be
evaluated easily through a recursive procedure (Supplementary Appendix
1).

\section{Phase-type semi-Markov models}\label{sec-smm}

A \emph{semi-Markov} model is defined by assuming that the sojourn
distribution for some state follows a distribution other than the
exponential, so that the rate of transition out of the state depends on
the length of time spent in the state. Hence the ``history''
\(\mathcal{F}_t\) consists of the time since the individual last moved
to the state they are in at time \(t\), so the intensities can be
written as \(q_{rs}(t)\) if \(t\) is redefined as the time since last
entry to state \(r\).

\begin{figure}

\centering{

\pandocbounded{\includegraphics[keepaspectratio]{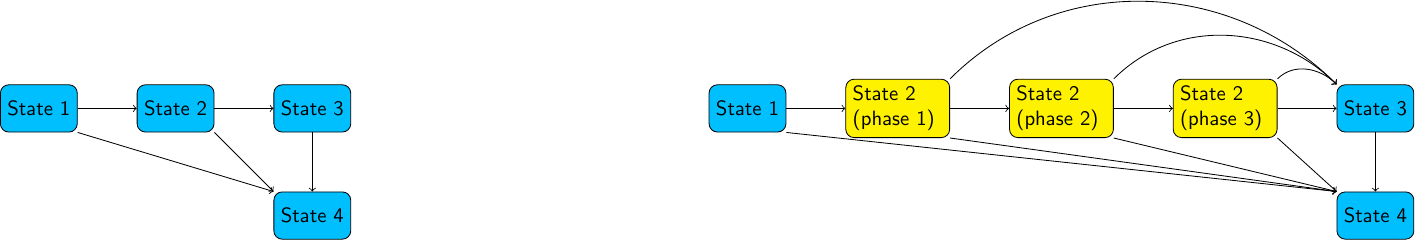}}

}

\caption{\label{fig-phasetype}An example of a multi-state model where
one state has a phase-type sojourn distribution with three phases. Left:
observable state space. Right: latent state space.}

\end{figure}

In \textcite{titman2010model}, a \emph{Coxian phase-type} distribution
is used as the sojourn distribution. An (observable) state \(r\) with
this sojourn distribution is replaced by a sequence of latent states
\(r_{1},\ldots,r_{n(r)}\), known as \emph{phases}, where a larger number
of phases \(n(r)\) gives a more flexible distribution. An example is
illustrated in Figure~\ref{fig-phasetype}, where observable state
\(r=2\) is given a \(n(2)=3\)-phase distribution. If an individual
transitions to an observable state with a phase-type distribution, they
start in the first phase, \(r_1\). Their subsequent progression is then
governed by a Markov model on the latent state space, so that the
sojourn time in each phase is exponentially-distributed. From each phase
\(r_i\), instantaneous transitions are allowed either to the next phase
in the sequence \(r_{i+1}\) (with intensity \(q_{r_i,r_{i+1}}\)), or to
an ``exit'' from the phase sequence, to any one of the next potential
observable states (say \(s_1,s_2,\ldots\)), with intensity
\(q_{r_i,r_{exit}}\). For example, in Figure~\ref{fig-phasetype}, on
exiting the phase sequence for observable state \(r=2\), an individual
can transition to either state \(s_1=3\) or state \(s_2=4\). If such an
exit occurs, this is to state \(s_j\) with a constant probability
\(p_{r,s_j}\) (as in \textcite{titman2010model}, assumed to be
independent of the time spent in state \(r\)). The transition intensity
from \(r_i\) to \(s_j\) is then

\begin{equation}\protect\phantomsection\label{eq-phase-exit}{q_{r_i,s_j} = q_{r_i,r_{exit}} p_{r, s_j}}\end{equation}

A multi-state model with a phase-type sojourn distribution is an example
of a hidden Markov model, with transition intensities defined on the
latent state space, and known outcome probabilities: the probability of
observed state \(r\) given latent state \(k\) is 1 if \(k\) is one of
the phases of state \(r\), and 0 otherwise. The likelihood can therefore
be evaluated easily using the ``forward algorithm'', as detailed in
Supplementary Appendix 1.

\subsection{Constructing phase-type families inspired by standard
distributions}\label{sec-phaseapprox}

By increasing the number of phases, any time-to-event distribution can
be represented in theory (\textcite{xiangbin2025phase}). However, with
only intermittent observations of the state, there is a limit to how
well the distribution of the time between state transitions can be
characterised in practice. This time is interval-censored at best, and
at worst, some sojourns may not be observed at all (e.g.~periods spent
infected, in a two state model of infection and recovery, if individuals
are intermittently tested for the infection). Even two-phase sojourn
distributions require three parameters (one ``progression'' rate and two
``exit'' rates). The applied interpretation of the phase transition
rates is also unclear. Therefore, \textcite{titman2014estimating}
developed a model based on familiar two-parameter time-to-event
distributions, exploiting a phase-type approximation to enable the
likelihood to be calculated. In this approach, the sojourn distribution
for a ``semi-Markov'' state is defined as a two-parameter distribution
with shape \(a\), scale \(b\), and probability distribution function

\[ F(t | a, b) = F_p(t | \boldsymbol{\lambda}= {\mathbf{h}}(a) / b) \]

\(F_p\) represents a phase-type sojourn distribution with transition
intensities collected in the vector \(\boldsymbol{\lambda}\), and
defined as a function \({\mathbf{h}}()\) of the shape \(a\), scaled by
\(b\). \({\mathbf{h}}()\) is determined so that for any \((a,b)\), the
resulting phase-type distribution closely matches, say, a Weibull or a
Gamma distribution with shape \(a\) and scale \(b\). The function need
only be determined once, before observing any data, and stored in
software. The model can then be fitted directly to any data, without the
need to re-calculate \({\mathbf{h}}()\).

\textcite{titman2014estimating} used a numerical approach to obtain
\({\mathbf{h}}()\). For a particular shape \(a\), this can be found by
finding the rates \(\boldsymbol{\lambda}={\mathbf{h}}(a)\) that minimise
the Kullback-Leibler discrepancy \(KL(\boldsymbol{\lambda})\) between
the approximating phase-type model and the target Weibull or Gamma. To
obtain the approximation for all possible \(a\), \({\mathbf{h}}()\) was
defined as a spline function of \(a\) for each component of
\(\boldsymbol{\lambda}\), and the spline coefficients were determined by
optimisation. In practice, this was a challenging optimisation problem,
involving ad-hoc choices of spline knots to obtain the appropriate
flexibility, and tuning of the numerical integration (over a wide range
of times \(t\)) required to estimate \(KL(\boldsymbol{\lambda})\).

Here, instead, we obtain \({\mathbf{h}}()\) using a new, fast analytic
approach, which is easier to implement in software.
\textcite{bobbio2005matching} showed that given values for the mean,
variance and third moment, a phase-type sojourn distribution of the form
in Figure~\ref{fig-bobbio} can be found with those moments, as long as
the moments are within a set of bounds that depends on the number of
phases \(n\). As \(n\) increases, the bounds become wider, so a wider
range of instances of the target distribution can be represented.
Therefore, given a Gamma or Weibull distribution with a particular shape
and scale, we can calculate the corresponding moments, and deduce the
phase-type distribution with those moments. The phase-type parameters
are moderately simple closed-form functions of the moments.
Supplementary Appendix 2 gives details of the approximation formula and
required bounds.

\begin{figure}

\begin{minipage}[t]{\linewidth}

\raisebox{-\height}{

\pandocbounded{\includegraphics[keepaspectratio]{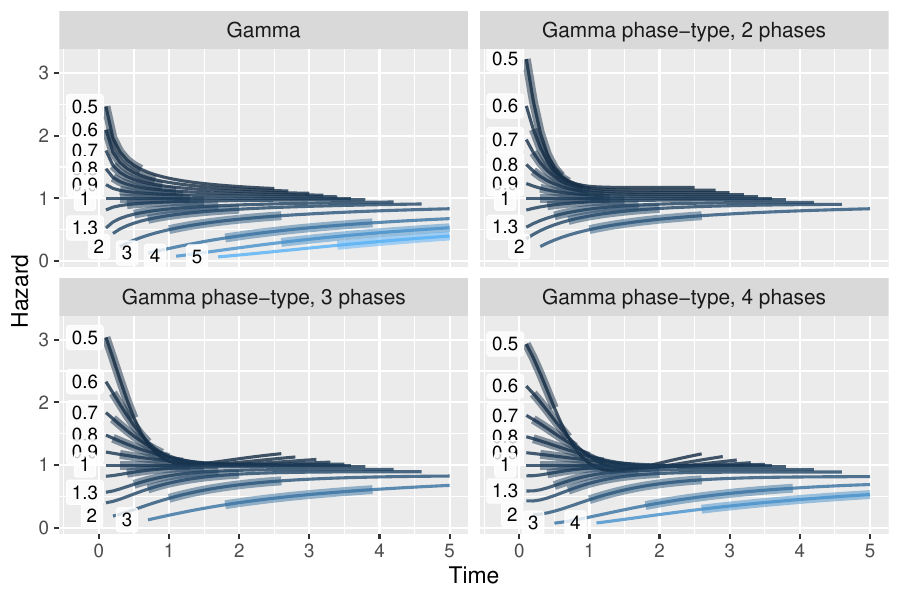}}

}

\end{minipage}%
\newline
\begin{minipage}[t]{\linewidth}

\raisebox{-\height}{

\pandocbounded{\includegraphics[keepaspectratio]{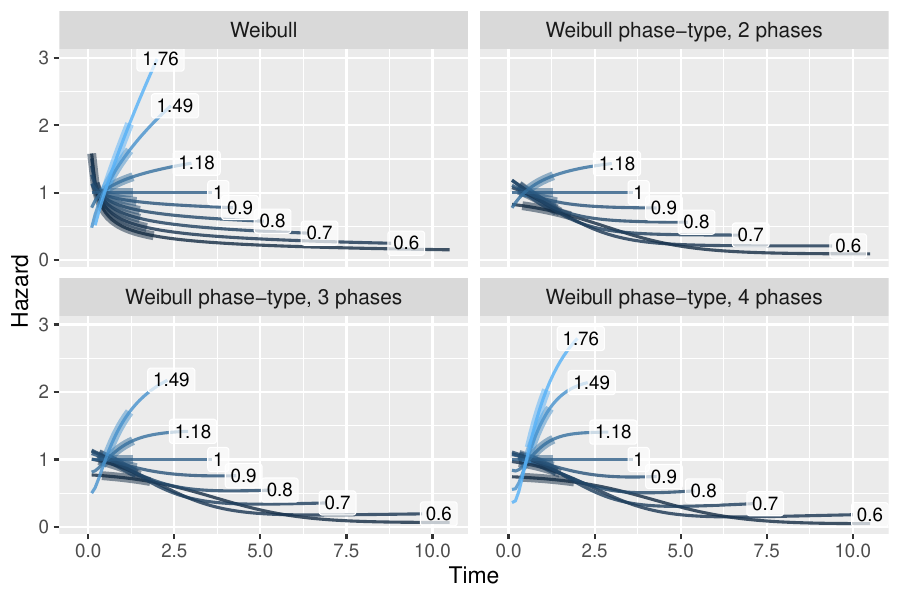}}

}

\end{minipage}%

\caption{\label{fig-hazard}Range of hazard functions for the Gamma and
Weibull-inspired phase-type models with 5 phases, compared to the
classical Gamma and Weibull, all with scale parameter 1 but different
shape parameters (indicated in the line labels). The displayed lines
extend as far as the 95\% probability ranges for each distribution,
while the interquartile ranges are shown as thicker lines.}

\end{figure}

\begin{figure}

\centering{

\pandocbounded{\includegraphics[keepaspectratio]{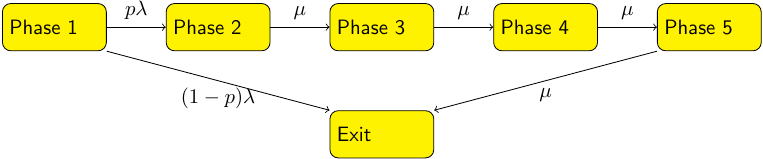}}

}

\caption{\label{fig-bobbio}Coxian phase-type model used to approximate
standard distributions by moment matching, illustrated for 5 phases. The
phase-type distribution is the distribution of the time from entering
phase 1 to entering the ``Exit'' state, in a continuous-time Markov
model with the indicated transition intensities. The particular
phase-type structure shown here is equivalent to the form in Bobbio et
al.~(2005), a mixture of two component distributions: (1) an
Exponential\((\lambda)\) with probability \(1-p\), and (2) a sum of an
Erlang\((n-1,\mu)\) and an Exponential\((\lambda)\) with probability
\(p\). The Erlang\((n,\mu)\) is a sum of \(n\) independent exponentials
with rate \(\mu\).}

\end{figure}

The hazard functions for the matching phase-type families with 5 phases
are compared with the standard Weibull and Gamma distributions in
Figure~\ref{fig-hazard} . For shapes greater than 1, the phase-type
families represent increasing hazards in a similar fashion to the
standard distributions. For shapes less than 1, all these phase-type
distributions have decreasing hazards in the bulk of the distribution,
but for shapes around 0.7 and less the hazard is U-shaped with an
increase in the upper tail, where the standard distributions are
monotonic throughout. The standard Weibull also has sharp decreases at
early times that are not matched by the corresponding phase-type model.
This pattern is the same for different numbers of phases. Note in
particular that increasing the number of phases does not give a closer
match to the standard model --- it merely increases the maximum shape
parameter that can be matched, resulting in families that are slightly
more flexible at representing increasing hazards. Shapes down to 0 are
supported in theory with any number of phases, however, low shapes give
extremely skewed distributions.

We emphasise here that for statistical modelling, we are not concerned
with matching the Weibull and Gamma exactly, since we do not believe
that real data are truly generated from any of these distributions.
Instead the goal is to obtain phase-type distribution families that are
useful in themselves for semi-Markov modelling, with key properties in
common with the distributions that they are inspired by.

\subsection{Using approximated phase-type sojourn distributions in
multi-state models}\label{sec-covariates}

To use the approximated sojourn distribution for a state \(r\) in a full
multi-state model, further extensions are needed. Recall that if there
is more than one state that an individual can move to on exit from state
\(r\) with a phase-type distribution, parameters \(p_{r,s_j}\) describe
the probability that the next state is \(s_j\), assumed to not depend on
the length of time spent in state \(r\). We extend the models described
in \textcite{titman2014estimating} to allow covariates to influence both
(a) the sojourn distribution, and (b) the next-state probability.

\textbf{Covariates on the sojourn time} Recall that the sojourn
distribution is defined by a set of transition intensities on the latent
space, \(\boldsymbol{\lambda}= {\mathbf{h}}(a) / b\). Replacing the
scale parameter \(b\) by \(b_0 \exp(-\boldsymbol{\beta}{\mathbf{x}})\)
(or reparameterising via a ``rate'' \(\rho=1/b\) and defining
\(\boldsymbol{\lambda}= \rho_0 \exp(\boldsymbol{\beta}{\mathbf{x}})\)),
defines an \emph{accelerated failure time} (AFT) model for the effect of
covariates \({\mathbf{x}}\) on the sojourn time. In an AFT model, if
\(P(T<t)\) is the CDF for scale \(b=1\), the CDF for scale \(b \neq 1\)
can be produced by scaling time, as \(P(T < t/b)\). This holds here,
because the phase-type distribution has a CDF of the form
\(P(T < t) = 1 - \boldsymbol{\alpha}
\exp(St) I\), where \(S\) is a matrix comprising the transition
intensities on the space of phases (see, e.g.
\textcite{cumani1982canonical}). Hence the ``time scaling'' can be
achieved by scaling all of the transition intensities by the same
scaling factor. Specifically, the progression and exit intensities
(rates) between phases of a state \(r\) are related to \({\mathbf{x}}\)
through the same log hazard ratio \(\boldsymbol{\beta}_r\):
\begin{align}
q_{r_i,r_{i+1}}  & = q_{r_i,r_{i+1}}^{(0)}  \exp(\boldsymbol{\beta}_r {\mathbf{x}}) \nonumber\\
q_{r_i,r_{exit}} & = q_{r_i,r_{exit}}^{(0)} \exp(\boldsymbol{\beta}_r {\mathbf{x}}) \label{eq-scale-covs}
\end{align}

\textbf{Covariates on next-state probabilities} Including covariates
only on the sojourn time defines a model where a covariate has an equal
effect on the rate of transition \(q_{r_i,s_j}\) to all exit states
\(j\) (substituting Equation \ref{eq-scale-covs} into
Equation~\ref{eq-phase-exit}). To relax this restriction, we also model
the next-state probability \(p_{r, s_j}({\mathbf{x}})\) in terms of
covariates, through a multinomial logistic regression,

\begin{equation}\protect\phantomsection\label{eq-next-covs}{ \log({p_{r, s_j}({\mathbf{x}})}/{p_{r,s_1}({\mathbf{x}})}) = \alpha + \boldsymbol{\gamma}_{rj} {\mathbf{x}}}\end{equation}

so that \(\boldsymbol{\gamma}_{rj}\) is the log odds ratio of transition
to state \(s_j\) for a unit increase in \({\mathbf{x}}\), where the odds
are respect to the first competing destination \(s_1\), and
\(1 / (1 + \exp(\alpha)) = p_{r,s_1} = p_{rs}\), the ``baseline''
next-state probability (dropping the ``1'' subscript for clarity).

Then if covariates are also included on the sojourn time through
Equation \ref{eq-scale-covs}, the full model for the transition
intensities on the latent Markov space (Equation~\ref{eq-phase-exit})
can be expressed as \begin{align}
q_{r_i,s_1}({\mathbf{x}}) & = q_{r_i,r_{exit}}^{(0)} \exp(\boldsymbol{\beta}_r {\mathbf{x}}) p_{rs_1}\\
q_{r_i,s_j}({\mathbf{x}}) & = q_{r_i,r_{exit}}^{(0)} \exp((\boldsymbol{\beta}_r + \boldsymbol{\gamma}_{rj}) {\mathbf{x}}) p_{rs_j}
\end{align}

This gives another interpretation of \(\boldsymbol{\gamma}_{rj}\), as
the relative hazard of transition to state \(s_j\), relative to \(s_1\),
from any of the latent phases \(i\) of state \(r\). If
\(\boldsymbol{\beta}_r>0\) but all \(\boldsymbol{\gamma}_{rj} = 0\),
covariates affect the sojourn time in state \(r\), but do not affect
which state happens next.

\section{Full likelihood and computation}\label{sec-comp}

Consider a multi-state model with a general state-transition structure,
and a state space \(\mathcal{S}\) partitioned into
\((\mathcal{S}_M, \mathcal{S}_P)\), where:

\begin{itemize}
\item
  States \(r \in \mathcal{S}_M\) are ``Markov'', in the sense of having
  outward transition intensities \(q_{rs}\) that are constant, or depend
  only on covariates \({\mathbf{x}}(t)\) that are piecewise-constant
  with respect to the time \(t\) since the start of the process, with
  \(q_{rs}({\mathbf{x}}(t)) = q_{rs}^{(0)}\exp(\boldsymbol{\beta}^T {\mathbf{x}}(t))\).
  The parameters for these states are then
  \({\boldsymbol{\theta}}_M = \{(q_{rs},\beta_{rs}): r \in \mathcal{S}_M, s \in \mathcal{S}\}\),
  excluding any parameters for which the instantaneous \(r-s\)
  transition is disallowed or a covariate has no effect.
\item
  States \(r \in \mathcal{S}_P\) are ``semi-Markov'', with phase-type
  shape-scale sojourn distributions, with parameters
  \({\boldsymbol{\theta}}_P = \{a_r, b_r, p_{rs}, \boldsymbol{\beta}_r, \boldsymbol{\gamma}_{rs}: r \in \mathcal{S}_P\}\)
  comprising shape parameters \(a_r\), baseline (i.e.~for covariate
  values of zero) scale parameters \(b_r\), baseline next-state
  probabilites \(p_{rs}\) (where permitted), and any effects of
  covariates on the scale (\(\boldsymbol{\beta}\)) or on next-state
  probabilities (\(\boldsymbol{\gamma}\)).
\end{itemize}

The full likelihood function comprises parameters
\({\boldsymbol{\theta}}= \{{\boldsymbol{\theta}}_M, {\boldsymbol{\theta}}_P\}\).
The values of the parameters in \({\boldsymbol{\theta}}_P\) for a
particular state \(r\) define transition intensities \(\lambda_{rs}\)
between the latent phases defining state \(r\). Therefore setting
\({\boldsymbol{\theta}}\) (and any covariate values) defines the
intensities of a Markov model on an expanded state space including the
observable states \(\mathcal{S}_M\) and all latent phases. The
likelihood can then be evaluated using the forward algorithm, described
in Supplementary Appendix 1.

The likelihood might either be maximised or used for Bayesian inference.
Both approaches are implemented in a new R package, \texttt{msmbayes},
available from \url{https://chjackson.github.io/msmbayes}. This wraps
around the \texttt{rstan} \autocite{rstan} interface to the Stan
\autocite{stan} software, which implements a quasi-Newton optimisation
algorithm (L-BFGS), or the ``no U-turn sampler'' Hamiltonian Monte Carlo
(HMC) algorithm for Bayesian inference. If the optimisation procedure is
used with improper uniform priors, this is equivalent to maximum
likelihood estimation, where the maximum likelihood is the posterior
mode. After optimisation, to produce interval estimates, a sample can be
drawn from the multivariate normal distribution defined by the estimates
and covariance matrix (obtained from the Hessian), giving a Laplace
approximation of the posterior.

Each shape parameter \(a_r\) is restricted during estimation to its
valid range for the moment-based phase-type approximation, which depends
on the number of approximating phases \(n\). In the Bayesian approach,
normal priors are assumed for all parameters (after transformation to an
unrestricted range if needed), or truncated normal distributions for the
\(a_r\). Either a Weibull or Gamma distribution may be used for the
phase-type approximation in each state.

A model can be fitted with a single R command, rather than requiring the
user to write Stan code. Any state transition structure can be used,
with different covariate models used for different parameters via
standard R linear modelling syntax. Any prior parameters can be
supplied, and posteriors summarised and processed easily via the
\texttt{posterior} (\textcite{posterior}) and \texttt{tidybayes}
(\textcite{tidybayes}) R packages. Vignettes giving worked examples are
available on the package website.

\section{Simulation-based calibration}\label{sec-sbc}

The correctness of the computational procedures and software is assessed
using simulation-based calibration \autocite{talts2018validating}. The
goals are to check that full MCMC (HMC) samples from the correct
posterior, to assess the relative speed and accuracy of the Laplace
approximation around the mode, and to compare the utility of the
Bayesian approach with current standard tools for maximum likelihood
estimation of multi-state models.

A range of datasets are simulated from the prior predictive distribution
for a given prior and model specification. The idea is that if the
computation produces the correct posterior for the given prior and data,
the posteriors produced by fitting the model to the simulated datasets
should span a range similar to the prior. Formally, the posterior
\(p({\boldsymbol{\theta}}| \mathbf{y})\), integrated over data
\(\mathbf{y}\) generated from the prior predictive distribution
\(p(\mathbf{y})\), is equivalent to the prior
\(p({\boldsymbol{\theta}})\). To check this, for each
\(i = 1, \ldots, N\), a value \(\tilde{\boldsymbol{\theta}}_i\) is
generated from the prior, hence a dataset \(\tilde{\mathbf{y}}_i\). Then
for each \(i\), a sample is drawn from the posterior distribution of
some estimand \(h({\boldsymbol{\theta}})\), and the rank \(r_i\) of
\(h(\tilde{\boldsymbol{\theta}}_i)\) among this sample is determined.
The resulting ranks \(r_1,\ldots,r_N\) should then have a uniform
distribution (\textcite{talts2018validating}). If the test fails, this
is evidence that either the algorithm to obtain the posterior is
inaccurate, or there are bugs in its software implementation.

The model and prior tested here are motivated by a hypothetical
infectious disease. The states include no infection (state 1) and
infection (state 2). The time spent with the infection (assumed to be
the same as the time testing positive), and the time until the next
infection, may not be exponentially distributed. For example, the risk
of being infected may increase since the time since the previous
infection, due to waning immunity. Beyond this, we assume that the risk
of being infected does not depend on the number of previous infections.

In each synthetic dataset \(i\), there are 100 individuals, distributed
in a 2/2/3/3 ratio between four age groups (``0-60'', ``60-70'',
``70-80'', ``80+''), and a 6/4 ratio (evenly within age groups) between
men and women. Their history of infections and clearance of infections
is generated from a continuous-time Markov or semi-Markov model. Each
person is assumed to be tested for the infection once every month for 12
months, giving 1200 intermittent observations of the infection state at
discrete times.

The priors are designed to be weakly informative, conveying that the
mean length of an infection is expected to be around 14 days, but this
may be as much as 30 days (note this is not the distribution of
individual infection durations, but uncertainty about the mean).
Similarly, the mean time to the next infection, since clearance of the
last, is expected to be 6 months, but this may be as much as 18 months.
The sampling models assessed are:

\begin{enumerate}
\def\labelenumi{(\alph{enumi})}
\item
  a Markov model, with age-sex group as a categorical covariate (with 8
  levels) on the rate of infection and the rate of clearance. Priors for
  the baseline intensities are \(\log(q_{12}) \sim N(-1.8, 0.6^2)\) and
  \(\log(q_{21}) \sim N(0.8,0.4^2)\), which represent the above
  judgements about times between events. N(0,1) priors are used for each
  log hazard ratio, that cover a wide range of practically-plausible
  hazard ratios (0.13 to 7).
\item
  a semi-Markov model implemented with a 5-phase approximation to a
  Weibull distribution. A \(N(0,0.35^2)\) prior is used for the log
  shape parameter, truncated on the supported range of 0 to log(2.1)
  (implying a 95\% credible interval of around 0.5 to 2 for the shape),
  and a normal prior for the scale parameter as in (a). Age-sex group is
  a covariate on the scale parameter, with the same priors used for the
  scaling factor \(1/b\) as for the hazard ratio in the Markov model.
\item
  a model where the two-state structure is extended to three states with
  an absorbing ``death'' state, with death permitted from either
  uninfected or infected. This gives ``competing risks'' of recovery or
  death from infection. For the log odds of death, relative to
  transition to the other living state, a normal\((0, 2.3^2)\) prior is
  used, which gives a roughly uniform distribution for the probability
  that the next state is death. Age-sex group also modifies the
  next-state probability, with (log) standard normal priors for the
  relative rate parameter \(\gamma\).
\end{enumerate}

Each model is fitted by two methods: MCMC, or posterior mode
optimisation followed by Laplace approximation. 1000 simulation
replicates are used.

The computational stability of the Bayesian approach is also compared
against the current standard maximum likelihood procedure for
multi-state modelling, using the \texttt{msm} package. The Markov model
(a) is fitted directly by maximum likelihood. To compare semi-Markov
models, since the phase-type Weibull/Gamma approximation is unavailable
in \texttt{msm}, the closest comparable model to (b) is fitted, that is
a two-phase model where phase transition rates are estimated directly
together with other parameters, rather than through a shape-scale
approximation.

Supplementary Appendix 3 describes an additional simulation study based
on the same motivating example but with a frequentist rather than
Bayesian design, which shows that the novel model gives an improved
characterisation of the sojourn distribution compared to a standard
Markov model, even when the data are generated from a different
distribution family than the fitted model.

\subsection{Results}\label{results}

\paragraph{Accuracy of different Bayesian computation
methods}\label{accuracy-of-different-bayesian-computation-methods}

\begin{figure}

\centering{

\pandocbounded{\includegraphics[keepaspectratio]{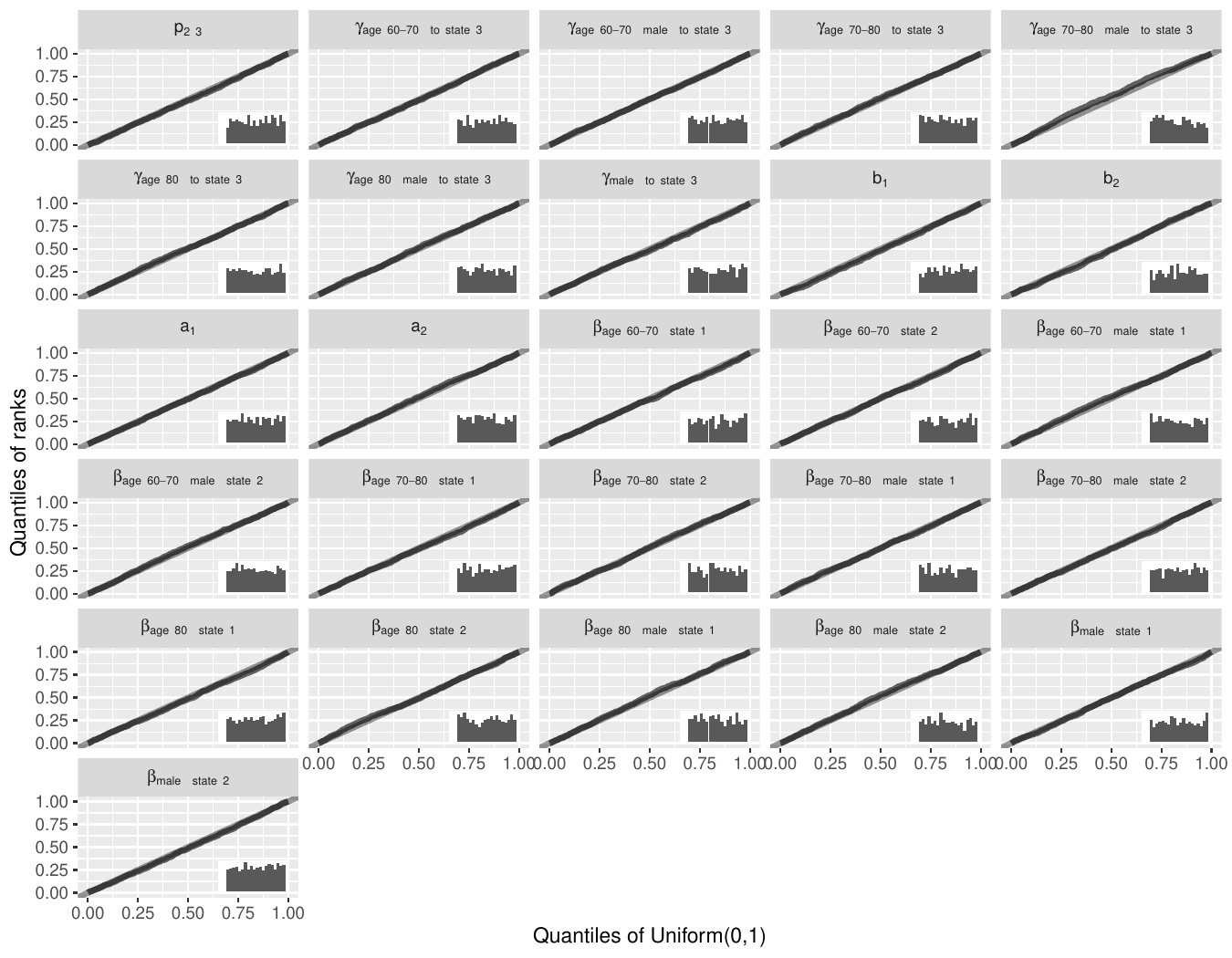}}

}

\caption{\label{fig-sbc-full-sample}Simulation-based calibration of a
phase-type shape/scale semi-Markov model with covariates, fitted by
MCMC. The distribution of the rank statistic for two parameters (log
transition intensities) over simulations is compared to a standard
uniform.}

\end{figure}

For the parameters of the full semi-Markov model (c),
Figure~\ref{fig-sbc-full-sample} illustrates that the ranks are
uniformly distributed, showing that the \texttt{msmbayes} package
accurately computes the posterior distribution in each model via the
MCMC algorithm from Stan. Similar results are shown in Supplementary
Appendix 4 for the simpler models (a) (Figures 1-2) and (b) (Figures
3-4).

\begin{figure}

\centering{

\pandocbounded{\includegraphics[keepaspectratio]{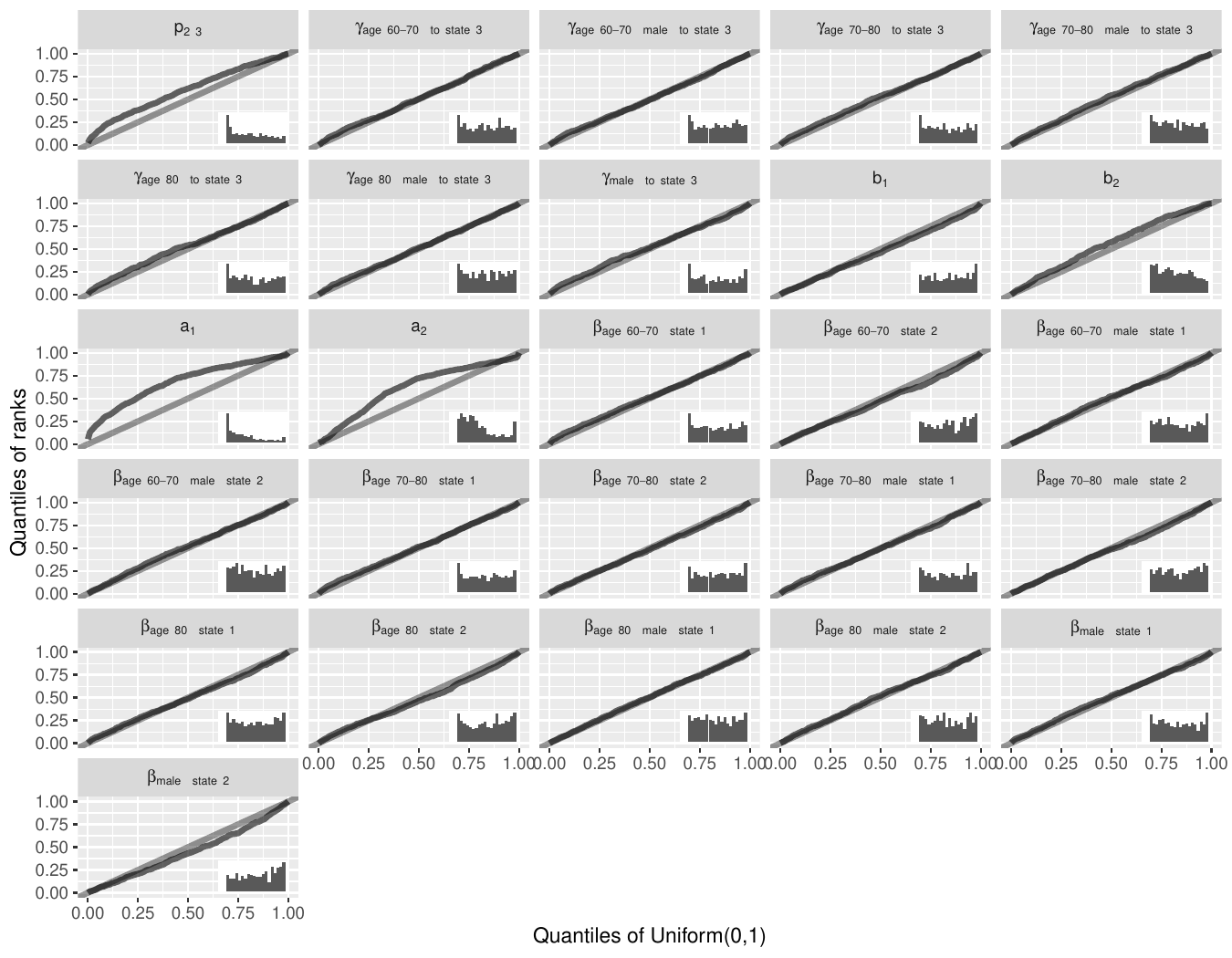}}

}

\caption{\label{fig-sbc-full-optimize}Simulation-based calibration of a
phase-type shape/scale semi-Markov model with covariates, fitted by
Laplace approximation around the posterior mode. The distribution of the
rank statistic for two parameters (log transition intensities) over
simulations is compared to a standard uniform.}

\end{figure}

For model (c), the Laplace approximation represents the true posterior
reasonably well (Figure~\ref{fig-sbc-full-optimize}), though there is
bias for some parameters, in particular the log shape parameters and log
odds of the next state. Taking the MCMC estimate as the true posterior,
the median log shape is generally overestimated by Laplace
approximation, with absolute bias having a median of 0.11 (95\% CI -0.07
to 0.35) over simulated datasets. The uncertainty about this parameter
is underestimated, with the posterior interquartile range having a
median bias of -0.19 (95\% CI 0.83 to 3.13). (Absolute rather than
relative biases are presented here because the true parameter value is
different in every simulation, being simulated from the prior, with a
prior credible interval of -0.7 to 0.6.) The Laplace approximation is
substantially faster, however, taking around 1\% of the run time of
MCMC. Note also that while it gives a biased median, the Laplace method
is based on an exact computation of the posterior mode.

Similar biases are seen for the shape parameter when using Laplace
approximation in the simpler semi-Markov model (b) (Supplementary
Appendix 4, Figures 3 and 4). For the Markov model (a) the biases from
using Laplace approximation are smaller (Supplementary Appendix 4,
Figures 1 and 2). For the parameter whose estimates appear the most
biased (Figure 2, \(\beta_{male,2,1}\)), the posterior median has an
absolute bias with median -0.06 (95\% CI -0.27 to 0.09), small in the
context of the prior credible interval of -2 to 2, and the interquartile
range has a bias with median -0.09 (-0.48 to 0.08).

\paragraph{Stability benefits of Bayesian estimation over maximum
likelihood}\label{stability-benefits-of-bayesian-estimation-over-maximum-likelihood}

\begin{figure}

\centering{

\pandocbounded{\includegraphics[keepaspectratio]{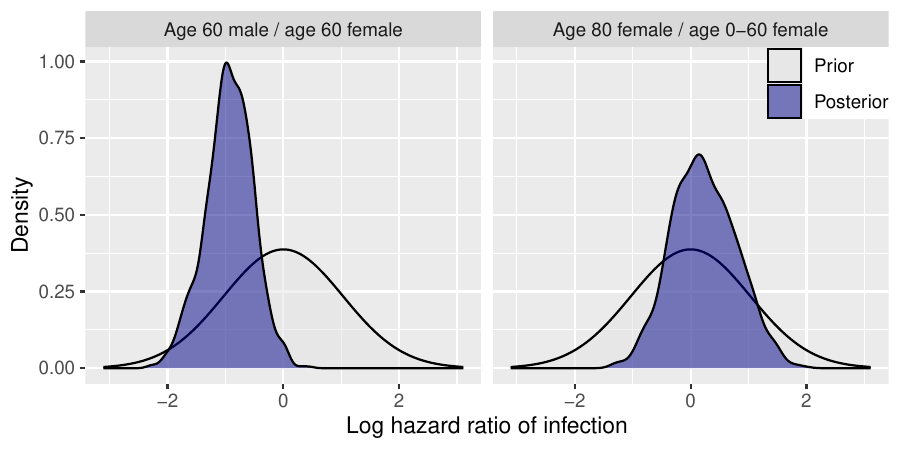}}

}

\caption{\label{fig-prior-post-cov}Prior and posterior distributions for
two selected log hazard ratios in the Markov model with covariates
fitted to the first simulated dataset}

\end{figure}

The results from fitting Markov model (a) illustrate the benefit of
Bayesian approaches in multi-state models that are weakly identifiable.
For 76\% of the simulated datasets, maximum likelihood estimation does
not converge to a maximum, with the Hessian deemed to be non-invertible
at the point where the optimisation algorithm terminated. This might be
expected after summarising the data, for example in the first dataset
there were only 10 infections from 8 individuals in the age 0-60, female
category. Even so, Bayesian estimation produces a confidently-converged
posterior, that is dominated by the weakly informative prior for the
covariate effects. Figure~\ref{fig-prior-post-cov} shows the posterior
is not much narrower than the prior for two selected hazard ratios in
the first simulated dataset, implying that the likelihood is near flat
in this dimension, but the data are informative in other dimensions.
Even if such a model cannot estimate these parameters from data, it
still provides a useful characterisation of uncertainty that can
motivate research to obtain specific further data.

\begin{figure}

\centering{

\pandocbounded{\includegraphics[keepaspectratio]{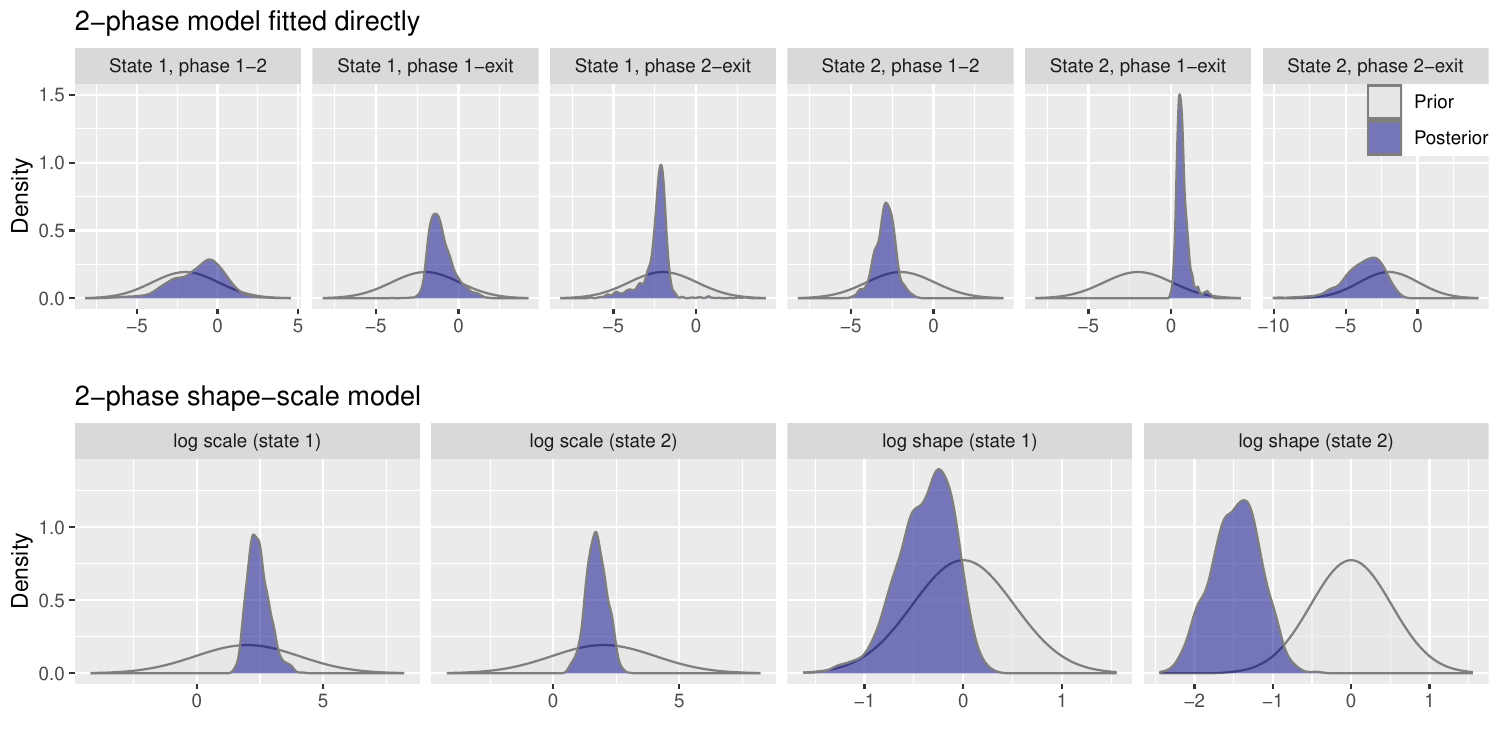}}

}

\caption{\label{fig-prior-post-nphase}Prior and posterior distributions
for parameters of a two-phase model with phase transition rates
estimated directly (top row) or via a Gamma shape-scale approximation,
for the first dataset in the simulation study}

\end{figure}

Maximum likelihood estimation of the two-phase model (using
\texttt{msm}) failed to converge for 91\% of simulated datasets, but
Bayesian estimation of semi-Markov model (b) succeeds for all. To
elucidate the problem with maximum likelihood, the two-phase model is
fitted to the first simulated dataset by Bayesian estimation, with
transition rates estimated directly using diffuse log normal\((-2,2^2)\)
priors, rather than through a shape-scale approximation.
Figure~\ref{fig-prior-post-nphase} shows the prior and posterior
distributions for the phase transition rates. For the phase 1-2 rate in
state 1, the posterior is similar to the prior, suggesting that there is
negligible information in the data for this parameter, so that maximum
likelihood estimation would fail. The comparable Gamma shape-scale
phase-type model has fewer parameters, that are better informed by the
data, hence the posteriors are further away from the priors. The
problems of eliciting prior information in phase-type models, and
comparison between fitted models, are discussed in more detail in the
context of an application in the following section.

\section{Application to cognitive function}\label{sec-app}

To illustrate the practical capability and computational scalability of
the method, we demonstrate how it might be used in a plausible
application. Code to reproduce all analyses in this section using
\texttt{msmbayes}, with a simulated dataset of the same structure, is
provided at
\url{https://chjackson.github.io/msmbayes/articles/cognitive.html}.

In the English Longitudinal Study of Ageing (ELSA) \autocite{elsa} a
cohort of people over 50 years old are surveyed around once every two
years. The survey included various measures of cognitive function. The
measure studied here is ``delayed word recall'', the number of words
recalled correctly from a 10-word list, after a delay spent answering
other survey questions. This outcome is categorised into one of four
states, as in \textcite{van2016multi}, who developed a Markov
multi-state model for the same dataset. The date of death is also
recorded. A random sample of 1000 people is selected from the full
dataset of 19802 people (giving a situation where Bayesian analysis
might be beneficial due to the greater influence of the prior for a
smaller sample size). The frequency of transitions between the states of
cognitive function and death, observed over intervals one time point to
the next, is illustrated in Table~\ref{tbl-statetable}. Note there are
only 67 deaths.

\begin{longtable}[]{@{}rrrrrr@{}}
\caption{Summary of transitions between states of cognitive function
(number of words recalled in a test) and death observed in the ELSA data
over intervals between successive
observations}\label{tbl-statetable}\tabularnewline
\toprule\noalign{}
& To: 1 & 2 & 3 & 4 & Death \\
\midrule\noalign{}
\endfirsthead
\toprule\noalign{}
& To: 1 & 2 & 3 & 4 & Death \\
\midrule\noalign{}
\endhead
\bottomrule\noalign{}
\endlastfoot
From: 1 (7-10 words) & 396 & 305 & 72 & 12 & 3 \\
2 (5-6 words) & 308 & 773 & 519 & 59 & 13 \\
3 (2-4 words) & 72 & 440 & 814 & 224 & 22 \\
4 (0-1 words) & 12 & 47 & 138 & 219 & 19 \\
\end{longtable}

\begin{figure}

\centering{

\pandocbounded{\includegraphics[keepaspectratio]{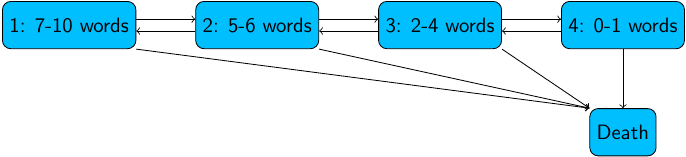}}

}

\caption{\label{fig-elsa-model}States and permitted continuous-time
transitions in the multi-state model for the ELSA data}

\end{figure}

A multi-state model is developed to describe the dynamics of change over
time in this cognitive measure, and how rates of death vary with the
cognitive state. Transitions in continuous time are permitted between
each adjacent cognitive state, and from each state to death
(Figure~\ref{fig-elsa-model}). Covariates include year of age (50-90,
median 60 at first visit), sex (54\% female) and highest level of
education (17\% tertiary, 47\% upper secondary, vs 36\% less). A linear
effect of year of age (since age 50) on each log transition rate is used
(as in \textcite{van2016multi}). In addition, the semi-Markov models
will allow an relaxation of the assumption that, given year of age, and
the current state, the rate of transition to the next state is constant.

The full model (with 50 parameters, from 10 transitions each with 4
covariates, excluding the semi-Markov extension) is not identifiable in
practice. If maximum likelihood is attempted (using \texttt{msm}), while
this appears to converge, it results in estimates and standard errors
for several parameters that are unrealistically large if interpreted as
degrees of belief, notably for the rates of death, for which the
transition counts are low.

To stabilise estimation, we can introduce prior information as part of a
Bayesian analysis. Strong priors for mortality rates, and effects of age
and sex on these, are obtained from published national mortality data.
Priors for other transition rates are assumed to be of a similar order
of magnitude, but with wide credible intervals. Priors for other
covariate effects are weakly informative, assuming (with 95\%
probability) that a single year of age, sex or education level will not
increase (or decrease) any transition rate by more than a factor of 7,
leading to a N(0, 1) prior for each \(\log(\beta_{rs})\).

The Markov model is compared with a semi-Markov model, in which the
sojourn distribution in each of the four living states is represented by
a 5-phase approximation to the Weibull or Gamma. Priors for the
parameters in the semi-Markov model that are compatible with the
substantive beliefs used in the Markov model are determined by
simulation.

Details of the priors and their derivation are given in Supplementary
Appendix 5.

\subsection{Computation}\label{computation}

The models are implemented by posterior mode optimisation and Laplace
approximation. Full MCMC, while feasible for the Markov models (with
around a day of computation) was not feasible for the semi-Markov
models. These computational demands come from the matrix exponential to
obtain the transition probabilities in the likelihood function:
\(P(t) = Exp(tQ({\mathbf{x}}))\) (Section~\ref{sec-models}). The matrix
exponential must be evaluated for each distinct \(Q({\mathbf{x}})\) and
time interval \(t\). If covariates are continuous, then each individual
in the dataset will typically have a different \({\mathbf{x}}\), as in
this dataset, where there are 767 distinct time intervals and covariate
values. The computational and memory requirements of the matrix
exponential also strongly depend on the number of rows or columns of the
matrix, which equals the number of (latent) states in the state space of
the (hidden) Markov model. While there are 5 states in the Markov model,
the semi-Markov model approximation here uses 5 latent phases to
represent each of the four living observable states, giving 21 latent
states overall.

\subsection{Comparison of model fit}\label{comparison-of-model-fit}

The overall goodness of fit of the models is compared informally via the
posterior mode, i.e.~the maximum penalised likelihood, by analogy with
AIC which prefers the model that minimises the minus log likelihood plus
the number of parameters. This neglects any influence of the priors,
which are intended to be of a similar strength between the three models.

The maximised log posterior is -4906 for the Markov model, -4871.2 for
the Gamma semi-Markov model and -4870.5 for the Weibull semi-Markov
model. The semi-Markov models therefore give a modest improvement in
fit, more than would be expected from the addition of only four extra
parameters (the shape parameters for the four living states).
Consistently with this improvement, the estimated shape parameters are
significantly less than one, indicating a decreasing hazard of with time
since state entry, for states 1, 3 and 4 (Table~\ref{tbl-soj-next}).

Note that formal cross-validation, or assessment of the \emph{absolute}
goodness of fit, would be challenging in general for this class of
models, due to the unbalanced intermittent observation
(\textcite{titman2010model}). Subsequent semi-Markov estimates presented
are from the Weibull model.

\subsection{Comparison of parameter
estimates}\label{comparison-of-parameter-estimates}

\begin{figure}

\centering{

\pandocbounded{\includegraphics[keepaspectratio]{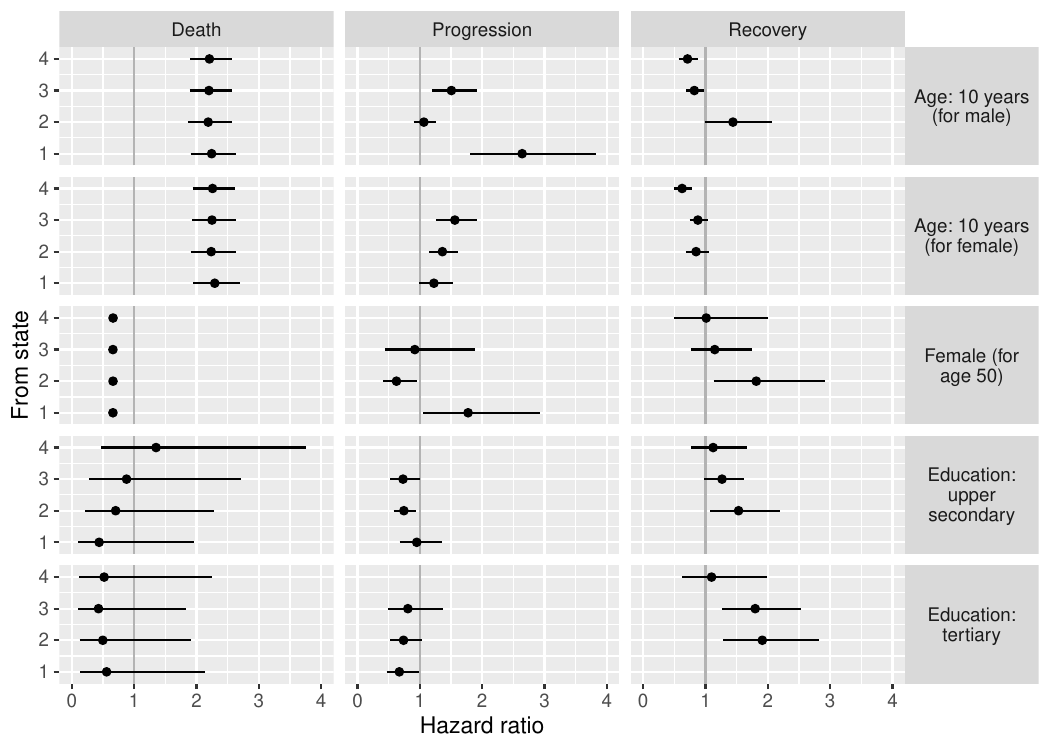}}

}

\caption{\label{fig-elsa-hrs}Covariate effects in the Markov model, as
hazard ratios modifying the transition rates from each of cognitive
states 1 to 4, to either progression (from states 1-3) to the next state
of severity, recovery to the previous state (from states 2-4), or death
(from any state). Posterior medians and 95\% credible intervals.}

\end{figure}

\begin{figure}

\centering{

\pandocbounded{\includegraphics[keepaspectratio]{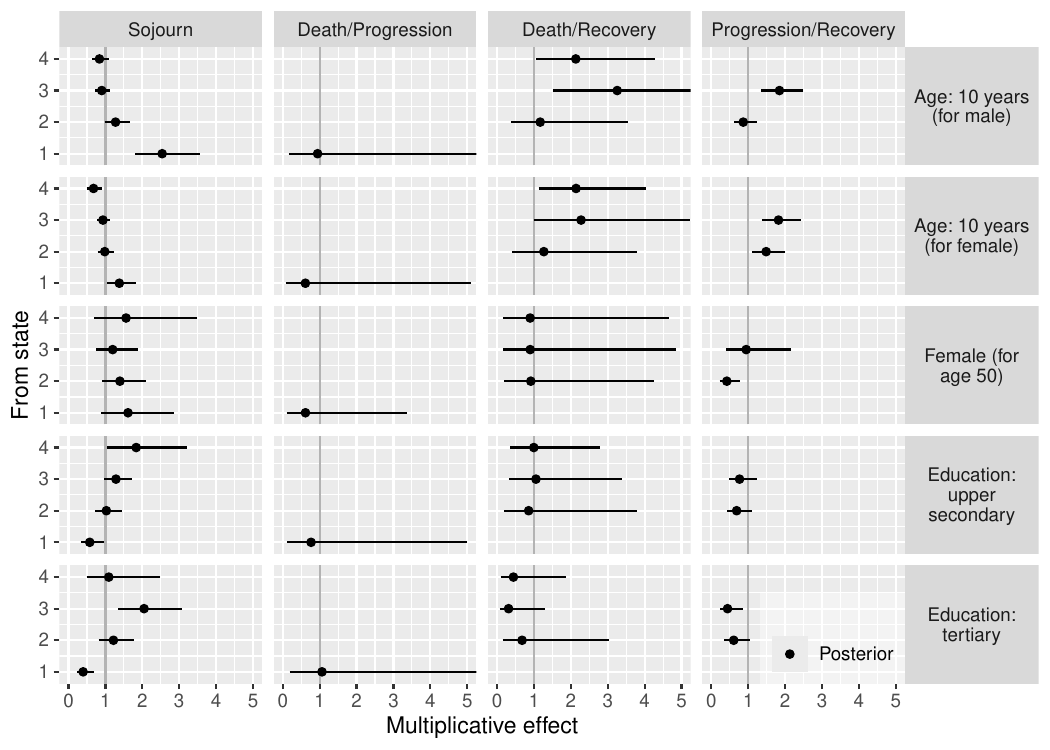}}

}

\caption{\label{fig-elsa-smm}Covariate effects in the semi-Markov model.
The first column ``Sojourn'' shows time acceleration factors modifying
the scale of the sojourn distribution (\(>1:\) faster time to event, or
higher risk). The remaining columns show multiplicative effects on the
relative risk of transition to a particular state, relative to some
``baseline'' state. For example, the first column ``Death/Progression''
shows the effect of the covariate on the relative risk of death compared
to progression to state 2, for people in state 1. Posterior medians and
95\% credible intervals.}

\end{figure}

Estimates of parameters (transformed to interpretable scales) are
presented in Table~\ref{tbl-soj-next}, including the mean sojourn time
and next-state probabilities for the baseline group (male, age 50,
lowest level of education). Transitions forwards and backwards between
the states of cognitive function are much more likely than death for the
youngest ages, but the probability of death increases with age.
Covariate effects on these are illustrated in Figure~\ref{fig-elsa-hrs}
for the Markov model and in Figure~\ref{fig-elsa-smm} for the
semi-Markov model.

Figure~\ref{fig-elsa-hrs} shows the effect of age on the hazard ratio of
death for both men and women, and the lower risk of death for women. Age
appears to increase the risk of progression, i.e.~decline in cognitive
function, and reduce the rate of recovery. The effect of sex on
transitions between cognitive states is unclear, while higher levels of
education are associated with lower risk of cognitive decline and higher
chance of recovery. The posterior for the effects of age and sex on
death is nearly identical to the prior from the national mortality data,
indicating the ELSA data provide little information about these. All
other covariate effects appear to be mainly informed by the data, since
the posterior credible intervals shown in this figure are much tighter
than the prior credible intervals of 1/7 to 7.

Figure~\ref{fig-elsa-smm} shows the effects of the same covariates under
the semi-Markov model, which are parameterised differently. Covariates
directly modify the sojourn time in a state through time acceleration
factors, and separately modify the relative risk of which state the
individual moves to next after this sojourn. While the message from the
effects on sojourn times is harder to see, the effects on next-state
probabilities are consistent with the Markov model. Age increases the
chance that when a person leaves a state, this is to death rather than
cognitive recovery (third column). Age also increases the chance of
progression in cognitive decline, relative to recovery. Higher levels of
education reduce the risk of progression, relative to recovery.

\subsection{Comparing model predictions between
subpopulations}\label{comparing-model-predictions-between-subpopulations}

A clearer illustration of the effect of covariates can be given, under
both models, by comparing the prediction of an \emph{absolute} outcome
between subgroups. As an example, we estimate the expected total length
of time spent alive in the states 1 and 2 (interpreted as mild cognitive
impairment or less), over a period of 10 years. This is defined as the
integral, over this period, of the probability of transition to any of
these states, for a person currently in state 1. Note that while, in
reality, age is time-dependent, the age covariate is assumed to be fixed
over this 10 year horizon (similarly to how time-dependent covariates
are fixed when computing the transition probabilities over intervals
between observations for the likelihood function, see
Section~\ref{sec-models}).

Comparisons between subgroups are done by fixing one covariate to
different values in turn, while standardising over the other covariates.
For example, to compare different ages, we define a standard population
whose distribution of sex and education roughly matches that of ELSA.
The posterior distributions of the predicted time are computed for each
member of the standard population while fixing age to 55, say, then
these distributions are averaged (by concatenating posterior samples) to
produce the standardised estimate for age 55. This is repeated for
different age values. Comparisons of genders and education categories
are done in a similar way.

These results quantify the greater time expected to be living free of
cognitive impairment for younger people, women, and people with higher
levels of education. The estimates agree between the Markov and
semi-Markov models, suggesting the conclusions are robust to the
assumptions made about the sojourn distribution.

\begin{table}

\caption{\label{tbl-soj-next}Mean sojourn times in state of cognitive
function, and next-state probabilities, under Markov and semi-Markov
models. From each of the four living cognitive function states, the next
state might be either progression (from states 1-3) to the next state of
severity, recovery to the previous state (from states 2-4), or death
(from any state). Posterior medians and 95\% credible intervals for a
man of age 50 years.}

\centering{

\fontsize{12.0pt}{14.4pt}\selectfont
\begin{tabular*}{\linewidth}{@{\extracolsep{\fill}}p{0.7in}p{0.7in}p{0.7in}p{0.7in}p{0.7in}p{0.7in}}
\toprule
State & Mean sojourn years & \multicolumn{3}{l}{Probability of next state}\\ 
  &  & Progression & Recovery & Death \\ 
\midrule\addlinespace[2.5pt]
\multicolumn{6}{l}{Markov model}\\
1  &  4.97 \newline {\footnotesize (3.12, 7.78)}  &  0.99 \newline {\footnotesize (0.97, 1.00)}  &    &  0.01 \newline {\footnotesize (0.00, 0.03)} & \\
2  &  1.76 \newline {\footnotesize (1.30, 2.39)}  &  0.80 \newline {\footnotesize (0.69, 0.88)}  &  0.20 \newline {\footnotesize (0.12, 0.31)}  &  0.00 \newline {\footnotesize (0.00, 0.01)} & \\
3  &  2.37 \newline {\footnotesize (1.72, 3.21)}  &  0.16 \newline {\footnotesize (0.09, 0.28)}  &  0.83 \newline {\footnotesize (0.72, 0.91)}  &  0.00 \newline {\footnotesize (0.00, 0.01)} & \\
4  &  1.14 \newline {\footnotesize (0.65, 1.96)}  &    &  1.00 \newline {\footnotesize (0.99, 1.00)}  &  0.00 \newline {\footnotesize (0.00, 0.01)} & \\
\hline
\multicolumn{5}{l}{Semi-Markov model} & Shape \\
1  &  2.97 \newline {\footnotesize (1.59, 5.48)}  &  1.00 \newline {\footnotesize (0.95, 1.00)}  &    &  0.00 \newline {\footnotesize (0.00, 0.05)}  &  0.64 \newline {\footnotesize (0.57, 0.72)}\\
2  &  1.65 \newline {\footnotesize (1.22, 2.20)}  &  0.69 \newline {\footnotesize (0.54, 0.81)}  &  0.31 \newline {\footnotesize (0.19, 0.46)}  &  0.00 \newline {\footnotesize (0.00, 0.02)}  &  1.08 \newline {\footnotesize (0.95, 1.21)}\\
3  &  3.14 \newline {\footnotesize (2.12, 4.79)}  &  0.16 \newline {\footnotesize (0.09, 0.26)}  &  0.84 \newline {\footnotesize (0.74, 0.91)}  &  0.00 \newline {\footnotesize (0.00, 0.01)}  &  0.75 \newline {\footnotesize (0.63, 0.87)}\\
4  &  1.95 \newline {\footnotesize (1.06, 3.66)}  &    &  0.99 \newline {\footnotesize (0.95, 1.00)}  &  0.01 \newline {\footnotesize (0.00, 0.05)}  &  0.73 \newline {\footnotesize (0.57, 0.89)}\\
\bottomrule
\end{tabular*}

}

\end{table}%

\begin{figure}

\centering{

\pandocbounded{\includegraphics[keepaspectratio]{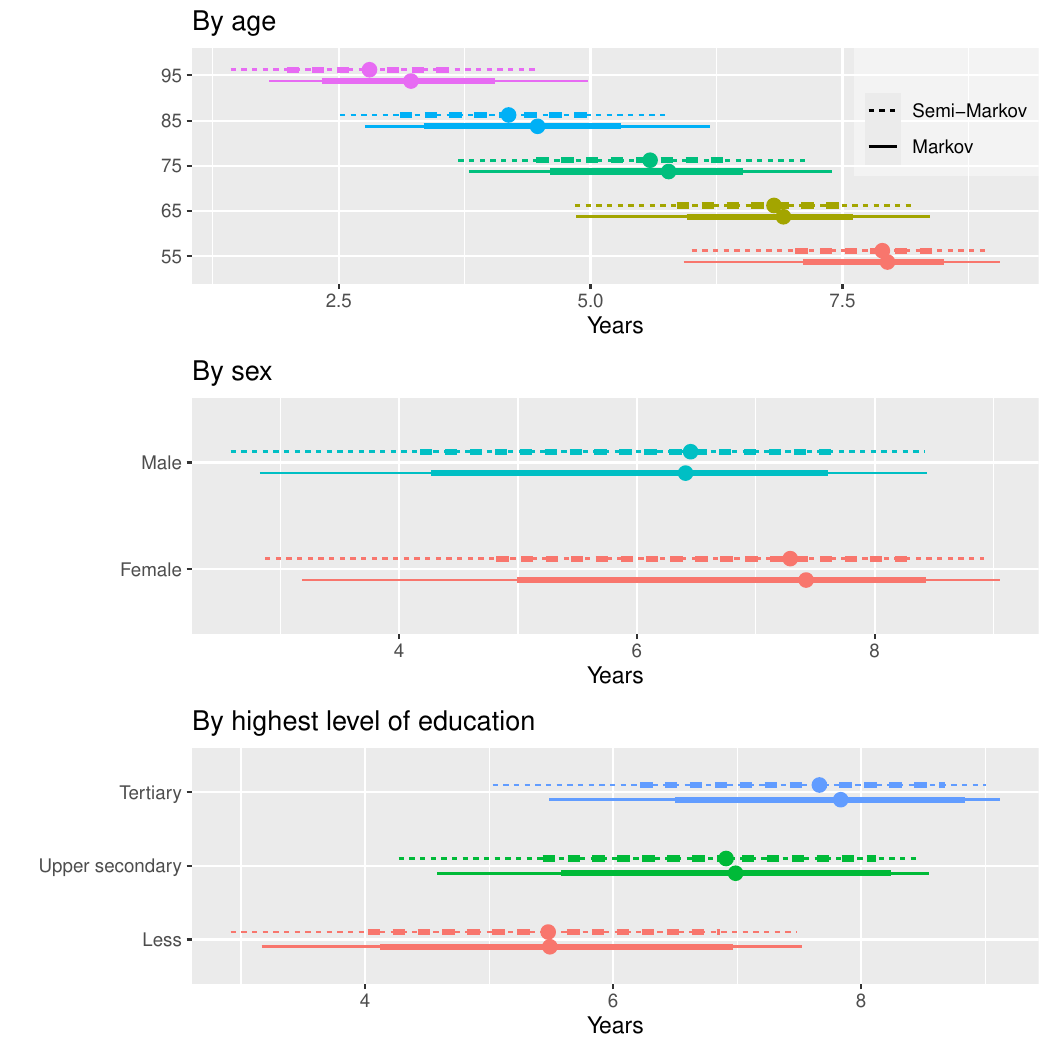}}

}

\caption{\label{fig-elsa-totlos}Predicted total time spent with no or
mild cognitive impairment over 10 years (states 1 and 2, recall of 5-10
out of 10 words.) Standardised comparisions between ages, between sexes
and between levels of education.}

\end{figure}%

\section{Discussion}\label{sec-discussion}

This paper has introduced a computational method that makes semi-Markov
multi-state modelling of intermittently-observed data practicable in
software. Analytic moment matching is used to easily construct families
of phase-type sojourn distributions that approximate the Weibull and
Gamma, which allows the likelihood to be evaluated as for a hidden
Markov model. The phase-type approximation method is extended to handle
covariates, and it is implemented in a general-purpose R package,
\texttt{msmbayes}, that fits Bayesian multi-state models with any
transition structure and covariates. The software is tested using
simulation-based calibration, and the illustrative application shows the
use of the methods in a typical applied workflow.

The approach has some limitations, however, in terms of flexibility and
scalability. The two-parameter shape-scale sojourn distributions used
here are more flexible than the exponential distributions used in these
models, but are still limited. Though in practice, we would expect
identifiability problems for more flexible distributions, particularly
if observations are infrequent relative to the sojourn time. Other
limitations of this class of semi-Markov models include independence of
the next-state probability and the sojourn time, and the assumption that
the entry time into the initial state is known --- potential extensions
are discussed by \textcite{titman2014estimating} (supplementary
material).

A stronger limitation, perhaps, is the computational scalability of the
phase-type approximation. As the state space is expanded to include more
latent states, the matrix exponentials (required to evaluate the
likelihood) become rapidly more demanding of computation and memory
resources. We have shown Laplace approximation can give a useful
approximation to the posterior for larger datasets and models where MCMC
is infeasible, and in practice, only two or three phases may be needed
to get a usefully flexible family of sojourn distributions. By
implementing the models using a probabilistic programming language
(Stan), we have made it easier to extend them in the future to more
complex situations (e.g.~hierarchical models, informative observation,
hidden Markov models with general outcomes), though problems of
identifiability and computational scalability would remain.

Model checking and comparison remains a challenge for parametric
multi-state models where transition times are unknown, in particular
where observations are not on a regular grid \autocite{titman2010model}.
It is increasingly common to assess statistical models using
cross-validation, but this is difficult in situations such as this where
the data structure is not exchangeable \autocite{vehtari2020cross}. An
alternative strategy is to compare estimates from a parametric model
with nonparametric estimates, which is possible for restricted classes
of multi-state models, e.g.~acyclic transition structures
\autocite{gu2023maximum} or with known transition times
\autocite{gomon2025nonparametric}.

For complex Bayesian models such as these, where parameters do not all
have simple interpretations, another difficulty is to obtain priors that
reflect substantive information. A general strategy is to simulate from
and check the prior predictive distribution, but the specific approaches
we took here (Supplementary Appendix 5) were ad hoc. There is an active
area of research about choosing prior parameters to minimise the
discrepancy between the prior predictive distribution and information
elicited about observable quantities
\autocite{manderson2023translating,bockting2024simulation}. Tools for
routine prior sensitivity analysis are also increasingly accessible
\autocite{kallioinen2024detecting}.

Note there are different forms of time-dependence in multistate models.
This paper deals with semi-Markov models, which express dependence of
the transition intensities on the time spent in the current state. This
is a different issue from dependence on time-varying covariates, or on
the time since the start of the multi-state process, which also requires
care if the data are intermittently-observed. When including covariates
in the model for the sojourn time and the model for the next state
(Section~\ref{sec-covariates}), we assumed the transition intensity is
piecewise-constant, so that the intensity at any time is assumed to
depend on the covariate values at the immediately-previous observation.
This allows the transition probabilities for the likelihood to be
obtained by matrix exponentiation. This may lead to bias, however, when
the covariate is smoothly-varying over time, and the observation
intervals are infrequent (\textcite{kendall2025beyond}). For example, in
the cognitive function application, intensities are assumed constant
over the (largely) two-year observation intervals, even though a
person's age is known to increase over these intervals. In practice, a
finer observation grid could be used if the covariate is known at all
times (as with age or time itself), while integrating over the
potentially-unknown state at grid points if necessary (e.g.~using the
\texttt{censor\_states} option in \texttt{msmbayes}). However, a more
realistic model would represent age or time as a flexible function, with
transition probabilities obtained by numerical ODE solving
(\textcite{kendall2025beyond}; \textcite{titman2011flexible}).

The phase-type approach to semi-Markov modelling has a particular
advantage (compared to explicit integration over unknown transition
times) when there are reversible transitions. In practice, multistate
models of this kind should pay heed to the mechanism represented by the
states and how the measurements are generated. In the cognitive function
example, much of the apparent reversibility in the ``word recall''
outcome may be due to noise in the measurement, rather than true
reversibility in cognitive function. While mild cognitive impairment is
often considered to be reversible (e.g. \textcite{sanz2022transition}),
dementia is a permanent condition. A hidden Markov model, with one or
more cognitive measurements generated conditionally on an underlying
cognition state, may be more suitable --- this approach was used by
\textcite{jackson2016modelling} in the context of psoriatic arthritis
with a similar reversible ``disease activity'' state. Multi-state models
are suitable in cases where the state itself is meaningful, and this
paper expands the range of multi-state models that can be used in
practice.

\section*{Acknowledgements}

This research was supported by the MRC, programme grant code
MC\_UU\_00040/4.

The English Longitudinal Study of Ageing was supported by the National
Institute on Aging, of the National Institutes of Health, under Award
Number R01AG017644, and NIHR Policy Research Programme (HEI)
198\_1074\_03. The content is solely the responsibility of the author
and does not necessarily represent the official views of these
organisations.

Thanks to Andrew Titman for providing more details of the phase-type
approximation in \textcite{titman2014estimating}, and to the two
reviewers and associate editor whose input strengthened the paper.

\section*{Appendices}

\begin{enumerate}
\def\labelenumi{\arabic{enumi}.}
\item
  Likelihood for a hidden Markov model
\item
  Details of moment-matching procedure for approximating standard
  time-to-event distributions by phase-type distributions
\item
  Frequentist simulation study
\item
  Additional plots for the simulation-based calibration analyis
\item
  Specification of prior distributions for semi-Markov models
\end{enumerate}

\section*{References}

\printbibliography[heading=none]

\end{document}